\documentstyle[prb,aps,twocolumn]{revtex}
\begin{document}
\draft
\title{Cyclotron resonance lineshape in a Wigner crystal}
\author{Peter Johansson\cite{Address}}
\address{NORDITA, Blegdamsvej 17, DK--2100 Copenhagen {\O}, Denmark}
\date{To appear in Phys. Rev. B
(Rapid Comm.) 15 November 1994.}
\maketitle
\begin{abstract}
The cyclotron resonance absorption spectrum
in a Wigner crystal is calculated.
Effects of spin-splitting are modelled by
substitutional disorder, and calculated in the
coherent potential approximation.
Due to the increasing strength of the dipole-dipole interaction,
the results show a crossover from a double-peak spectrum
at small filling factors to a single-peak spectrum
at filling factors $\agt 1/6$.
Radiation damping and magnetophonon scattering
can also influence the cyclotron resonance.
The results are in very good agreement with experiments.
\end{abstract}
\narrowtext

Two-dimensional (2D) electron systems
in strong magnetic fields
have been studied intensively for the last 15 years.
Most of the work has been concentrated on transport
properties.
Important information can also be extracted from measurements
of high-frequency properties such as
photoluminescence and the cyclotron resonance.
A large number of experimental and theoretical studies
have dealt with these phenomena.\cite{PL,CR,Bess,Summ}
The interactions with the low-energy degrees of freedom
influence the detailed structure of the high-energy
resonance.

Experiments on 2D electron systems at small
Landau level (LL) filling factors $\nu$
by Besson {\it et al}.\cite{Bess} and
Summers {\it et al}.\cite{Summ}
revealed intriguing changes
in the cyclotron resonance absorption spectra as
the filling factor $\nu$, and temperature were varied.
For very small $\nu$ the spectra show two peaks.
With increasing $\nu$, the absorption peak
at the lower frequency increases in relative strength,
and the peaks are shifted.
Eventually, for $\nu \agt$ 1/6 the two peaks merge.
The spectra also change in qualitatively different
ways  with increasing temperature for $\nu$
smaller and larger than $\approx$ 1/10, respectively.
It was speculated that this
signaled a phase transition, for example
the formation of a Wigner crystal.

In a recent paper, Cooper and Chalker\cite{Coop} introduced
a model that explains the experimental results
without invoking a phase transition.
The idea behind the model is that, due to band-structure
effects, spin-up and spin-down electrons have slightly
different cyclotron resonance frequencies.\cite{Hopk}
At a finite temperature there are electrons with both spins,
and the 2D system is a disordered mixture
of the two spin species.
In this model, the electrons (i.e.\ their guiding centers)
occupy the lattice sites of a Wigner crystal (WC).
The optical properties of this system
near the cyclotron resonance frequency
can be described in terms of excitations, ``excitons'',
from the lowest LL to the next LL.
The excitons
propagate on a triangular lattice with substitutional disorder.
The propagation is caused by the electromagnetic
dipole-dipole interaction,
and its strength relative to the
splitting between the spin-up and spin-down
cyclotron frequencies determines the shape of the
absorption spectrum.
In Ref.\ \onlinecite{Coop}, the averaging over different
disorder configurations was done numerically.

This paper presents a study of the effects of spin-splitting
disorder treated within the coherent potential
approximation (CPA),\cite{Rick} which makes it possible
to obtain results that agree with experiment,
with only a small calculational effort.
We also present a thorough discussion of the
physics behind the results.
Moreover, we deal with two other processes that
broaden the cyclotron resonance:
radiation damping and thermal
motion of the electron guiding centers treated in terms of
magnetophonons (MP's).\cite{MPs,PJ}
Radiation damping gives a contribution
to the resonance width comparable to
experimental values once the electron density
is large enough.
In the case of a double-peak spectrum,
the MP's broaden the high-frequency peak
more than the low-frequency peak.

The model that we will study describes excitons
propagating in a WC with substitutional
disorder.
The WC is a convenient model system, but the results
of the calculations should be similar also in
an electron liquid.
The Hamiltonian can be written
\begin{equation}
H=\sum_{i}\hbar \omega_{i}c_{i}^{\dag}c_{i} +
   \sum_{ij}t_{ij}c_{i}^{\dag}c_{j},
  \label{Ham}
\end{equation}
where $c_{i}$ annihilates an exciton at site $i$.
The site cyclotron frequency $\omega_{i}$
equals the bare cyclotron frequency
$\omega_{c}=eB/m^{*}$
($B$ is the magnetic field and $m^{*}$ the electron effective mass)
for the majority spins, and
$\omega_{c}+\delta\omega$ for the
minority spins.\cite{Hopk,Note}
The exciton-hopping matrix elements due to the dipole-dipole
interaction are given by
\begin{equation}
t_{ij}=
 -\frac{e^2l_{c}^{2}}{8\pi\epsilon_{0}\epsilon_{r}}
|{\bf R}_{i}-{\bf R}_{j}|^{-3},\ \ \
i\neq j,
\label{tij}
\end{equation}
while $t_{ii}=-\sum_{j,j\neq i}t_{ij}$.
As will be seen, this fixes the cyclotron frequency
at $eB/m^{*}$ in a WC without disorder,
in accordance with Kohn's theorem.\cite{Kohn}
In Eq.\ (\ref{tij}), ${\bf R}_{i(j)}$ are lattice vectors of
the WC,
$\epsilon_{0}$ is the
dielectric constant of vacuum, $\epsilon_r$ is the
relative dielectric constant of the material (GaAs),
and $l_{c}=(\hbar/eB)^{1/2}$ is the magnetic length.
The Fourier transform $t_{{\bf q}}$ of the hopping
matrix elements yields the exciton
dispersion in a system without disorder.
It is proportional to the trace
of the dynamical matrix of a Wigner crystal in the
absence of a magnetic field,\cite{MPs,PJ}
and can be related
to the transverse and longitudinal phonon frequencies
$\omega_{T{\bf q}}$ and $\omega_{L{\bf q}}$ by
\begin{equation}
t_{{\bf q}}=\hbar(\omega_{T{\bf q}}^2+\omega_{L{\bf q}}^2)/
(2\omega_c).
\label{tq}
\end{equation}
For large magnetic fields and small filling factors
Eq.\ (\ref{tq}) is, for all practical purposes, equivalent
to the {\em magnetoplasmon} dispersion relation
resulting from a lattice dynamics treatment.\cite{PJ}

The absorption probability of a circularly polarized
photon with frequency $\omega$ near the cyclotron frequency,
impinging on the electron system at right angle, is
\begin{equation}
P(\omega)=-\frac{e^{2} n_{e} \omega_{c} }
   {2m^{*}\epsilon_{0}\sqrt{\epsilon_{r}}c_{0}\omega}
   {\rm Im}\left[ G({\bf q}=0,\omega)\right].
\label{Pdef}
\end{equation}
This result
treats the ${\bf A}_{{\rm rad}}\cdot{\bf j}$ interaction
between the radiation and the electrons
to lowest order (Fermi golden rule).\cite{Saku}
In Eq.\ (\ref{Pdef}), $n_e$ is the electron density,
$c_0$ is  the speed of light in vacuum,
and $G({\bf q},\omega)$ is the
Fourier transform of the retarded exciton Green's function
\begin{equation}
G_{ij}(t)=-i\theta(t)\langle [c_{i}(t),c_{j}^{\dag}(0)]
\rangle.
\label{Gdef}
\end{equation}

We use the CPA to calculate $G({\bf q},\omega)$.
The CPA is known to give correct results in a
number of important limits,\cite{Rick}
and Persson and Ryberg\cite{BNJP}
used it successfully
to solve the formally equivalent problem of optical
absorption in an isotopically mixed adsorbate layer.
The disordered system,
where the exciton frequency at a site is $\omega_c$
with probability $(1-c)$
and $\omega_c+\delta\omega$ with probability $c$,\cite{Note2}
is replaced by a translationally invariant system
(effective medium) where the exciton
frequency is $\omega_c+\sigma(\omega)$ everywhere.
The exciton self energy $\sigma(\omega)$,
which must be determined self-consistently,
is frequency-dependent and, in general, complex.
The Green's function can be written
\begin{equation}
G({\bf q},\omega)=[\omega+i\gamma-\omega_c-\sigma(\omega)-
    t_{{\bf q}}/\hbar]^{-1},
\label{Gexpr}
\end{equation}
where $\gamma$ represents exciton damping due to other
mechanisms than disorder (see below).
In the CPA,  one real site
(with frequency $\omega_c$ or $\omega_c+\delta\omega$)
is placed in the effective medium of $\omega_{c}+\sigma(\omega)$
sites.
Requiring that the scattering off the real site, treated
to all orders, should vanish on the average
yields an equation for $\sigma(\omega)$,\cite{Rick}
\begin{equation}
\sigma(\omega)=\frac{c \delta\omega}
{1-[\delta\omega-\sigma(\omega)]G_{00}(\omega)}.
\label{CPAeq}
\end{equation}
The diagonal Green's function in
real space
\begin{equation}
G_{00}(\omega)=A_{{\rm BZ}}^{-1}
\int_{{\rm BZ}} d^2q G({\bf q},\omega),
\label{G00def}
\end{equation}
where $A_{{\rm BZ}}$ is the area of
the first Brillouin zone (BZ).
In the calculations, we have evaluated
Eq.\ (\ref{G00def}) accurately,
but the analytic approximation\cite{NoteW}
\begin{equation}
G_{00}(\omega)=\frac{\hbar}{W}\left[
\frac{\hbar\alpha}{W}
\ln\left(\frac{\hbar\alpha}{\hbar\alpha-W}\right)-1 \right]
+\frac{\hbar/2}{\hbar\alpha-W},
\label{Gsimp}
\end{equation}
where $\alpha=\omega+i\gamma-\omega_c-\sigma(\omega)$ and
$W$ is the exciton bandwidth, gives almost the same results.

There are a number of processes that can cause
exciton damping described by the constant $\gamma$
in Eq.\ (\ref{Gexpr}).
Radiative decay is one of the most important,
and it can be calculated
in a standard way.\cite{Saku}
Excitons with a wave vector larger than
$\sqrt{\epsilon_{r}}\omega_{c}/c_{0}$ do not radiate.
For smaller wave vectors the radiation damping is
\begin{equation}
\gamma_{{\bf q}}=
\frac{e^2 n_e}{4m^* \epsilon_0 \sqrt{\epsilon_r} c_0}
\left[ \frac{\omega_c}{\sqrt{\omega_c^2 - q^2 c_0^2 /\epsilon_r}}
+\frac{\sqrt{\omega_c^2 - q^2 c_0^2 /\epsilon_r}}{\omega_c}
\right]
\label{Radrate}
\end{equation}
(neglecting exciton dispersion).
This result resembles those for the decay of free
excitons in quantum wells.\cite{Decay}
Numerically
$\hbar\gamma_{{\bf  0}}$=0.013 meV
when $n_{e}=10^{11}$ cm$^{-2}$.

It should be emphasized that Eq.\ (\ref{Pdef}) yields $P\le 1$
with radiation damping included, whereas neglecting
it may give $P>1$.
On resonance $\omega=\omega_c$, and without
disorder $\sigma(\omega)=0$,
setting $\gamma=\gamma_{{\bf q}}$ in Eq.\ (\ref{Gexpr}),
and then using Eq.\ (\ref{Pdef}) yields $P=1$.
The present calculations suppress the
${\bf q}$ dependence of the damping. The constant
$\gamma=\gamma_{{\bf q}=0}+\gamma_{{\rm ph}}$,
using
$\hbar\gamma_{{\rm ph}}=0.005$ meV for the
phenomenological damping constant.

Figure \ref{Fig1} shows the main results of this calculation.
In Fig.\ \ref{Fig1}(a) the magnetic field and
temperature are kept constant, while the
electron density varies.
The qualitative behavior of the spectra as a function
of $n_{e}$ is in close agreement with experiment
(cf.\ Fig.\ 2 in Ref.\ \onlinecite{Summ}).
The bare cyclotron
resonance lies at $\hbar\omega_{c}=27.64$ meV and
the minority spin resonance is at 27.49 meV\@.
If $n_{e}$ is chosen small enough, the absorption
spectrum has two isolated resonances at these
frequencies.
However, all the spectra displayed in Fig.\ \ref{Fig1}(a)
are modified by the dipole-dipole interaction,
since it is comparable in strength to the difference in resonance
frequency between the two spin species.
The exciton bandwidth $W\sim n_{e}^{3/2}$
increases from 0.10 meV at $\nu=1/14.2$ to
0.67 meV for $\nu=1/4.1$.

The physics behind the results of Fig.\ \ref{Fig1}(a)
can be understood in the following way
(cf.\ Refs.\ \onlinecite{BNJP} and \onlinecite{Coop}):
The transition between the two lowest
LL's at a site can be considered as a local oscillator.
The high-frequency, majority oscillators respond in
phase with the incident external radiation field
when it has a frequency near the lower resonance.
Then, the total field acting on the {\em minority} oscillators
is enhanced due to anti-screening,
and the low-frequency peak grows in strength
with increasing $n_{e}$.
At the same time, it is shifted towards higher
frequencies because the majority-oscillator response
is larger there.
At the high-frequency peak, on the other hand,
the majority oscillators absorb energy from the external
field and thus have a phase lag of $\approx \pi/2$.
The minority oscillators now respond out of phase with the
total field and therefore {\em lead} the external
field.
Consequently, the minority oscillators effectively emit
radiation, and this reduces the strength of the
second peak in the absorption spectrum.

Through all the changes of the absorption spectrum
its first moment stays constant at
$\omega_{m}=\omega_{c}+c\delta\omega$.
In the end, at high densities, only one peak
centered at $\omega_{m}$ remains.
This behavior is characteristic of motional narrowing.
Once the hopping matrix elements become large enough,
exciton hopping becomes so frequent that the excitons
are not much affected by the disorder.
In this regime, the disorder contribution to the
linewidth behaves as $\sim\delta\omega^{2}/W$,
and decreases with increasing density.
In an impurity-scattering language, the factor $1/W$
comes from the density of final states.
The other contributions to the linewidth are
not influenced by this narrowing,
and in Fig.\ \ref{Fig1}(a) the linewidth starts to grow
at the highest densities due to radiation damping.
This result is also in agreement with experiment.\cite{Summ}

Figure \ref{Fig1}(b) shows absorption spectra
at different temperatures and minority spin
concentrations. At low enough temperature, there are only
majority spins, and consequently only one peak.
When the temperature is increased, this peak splits in
qualitatively different ways depending on the density.

Consider now the effects of electron
thermal and zero-point motion.
The calculation of the exciton-hopping matrix elements
in Eq.\ (\ref{tij}) assumes that the electrons
are fixed at the lattice sites.
In reality they perform  vibrations
so that the matrix elements $t_{ij}$ are
time-dependent.
To quantify this time-dependence, return to Eq.\ (\ref{tij})
and replace the lattice site coordinates
${\bf R}_{i(j)}$ by
${\bf R}_{i(j)}+{\bf u}_{i(j)}$, where
${\bf u}_{i(j)}$ is the electron displacement
that can be described in terms of MP's.\cite{PJ}
Then expand the resulting expression to linear
order in ${\bf u}_{i(j)}$ recovering the
static exciton-hopping matrix element
plus a new term, describing exciton-MP interaction
due to anharmonicity.
This exciton-MP
interaction Hamiltonian can be written
\begin{equation}
H_{{\rm ex-mp}}=\sum_{{\bf q},{\bf q}'}
\left [ M_{{\bf q}{\bf q}'}a_{{\bf q}-{\bf q}'}
c_{{\bf q}}^{\dag}c_{{\bf q}'}
+H.c. \right ],
\label{Hexmp}
\end{equation}
where $a_{{\bf q}}$ annihilates an MP
of wave vector ${{\bf q}}$.

The expression determining the matrix element
$M_{{\bf q}{\bf q}'}$
is lengthy, however,
$M_{{\bf q}{\bf q}'}=0$ whenever
${\bf q}$ or ${\bf q}'$ or their difference
vanishes.
Thus, in a system without disorder,
as a direct consequence of Kohn's theorem,
a ${\bf q}={\bf 0}$
exciton is not scattered by the MP's.
But the exciton-MP
interaction still
influences the cyclotron resonance lineshape
in the presence of disorder.
The disorder scatters a long-wavelength exciton
into an intermediate state with a finite ${\bf q}$,
from where the exciton can decay due to the interaction with
the MP's.

We include the effects of these processes on
the exciton spectrum through a two-step calculation.
First, the decay rate
of an exciton with wave vector ${\bf q}$,
caused by $H_{{\rm ex-mp}}$,
is calculated in a system without disorder.
The Fermi golden rule gives
\begin{eqnarray}
\Gamma_{{\bf q}}=\frac{2\pi}{\hbar}
\sum_{{\bf q}'}|M_{{\bf q}{\bf q}'}|^2
&&\Big[ (1+N_{{\bf q}-{\bf q}'})
\delta(t_{{\bf q}}-t_{{\bf q}'}-\hbar \omega_{{\bf q}-{\bf q}'})
\nonumber \\
&&+N_{{\bf q}-{\bf q}'}
\delta(t_{{\bf q}}-t_{{\bf q}'}+\hbar \omega_{{\bf q}-{\bf q}'})
\Big ],
\label{Gammaq}
\end{eqnarray}
where $\omega_{{\bf q}}$ is an MP frequency and
$N_{{\bf q}}$
is the thermal MP occupation.
Then $i\gamma$ is replaced
by $i(\gamma+\Gamma_{{\bf q}}/2)$ in Eq.\ (\ref{Gexpr}),
and the same calculations as before yield
the exciton self energy and Green's function.

Figure \ref{Fig2} shows the results of these calculations.
The most interesting behavior occurs for an intermediate
value of the electron density ($n_{e}=3.2\times10^{10}$ cm$^{-2}$).
The low-frequency peak is slightly broadened when MP scattering
is included, but this is a minute change.
The high-frequency peak, on the other hand, is substantially
broadened, and only a weak shoulder remains.
This seems to be in qualitative agreement with the
experimental results of Ref.\ \onlinecite{Bess}.
There (see Fig.\ 1),
for a similar filling factor, the high-frequency peak is
more temperature-sensitive than the one at lower frequency.
Similar results were also found in Ref.\ \onlinecite{Coop},
where these thermal effects were modelled by frozen structural
disorder.
At lower densities, when the spectrum approaches that of
two isolated resonances, both peaks are somewhat broadened
by the MP scattering.
No results for the motional narrowing regime are shown,
since the MP scattering has little influence on the spectrum there.

The above results can be explained by very much the
same reasoning as was used earlier.
In the intermediate density regime and at frequencies
near the first peak, the oscillators respond
more or less in phase with each other. The small changes
in the dipole-dipole interaction caused by the
MP's do not then perturb the spectrum appreciably.
For higher frequencies, near the second peak,
nearest-neighbor oscillators may be out of phase
with each other.
Changes in the coupling due
to the MP's are much more important in this case.

In conclusion, this paper has presented a model calculation
of the cyclotron resonance lineshape in a Wigner crystal.
The effects of spin-splitting of the
cyclotron resonance  has been treated within the CPA\@.
The resulting absorption spectrum has two relatively
independent peaks, one for each spin species,
at low electron densities.
When the density increases, more and more strength
is transferred to the low-frequency peak due to
anti-screening effects
until, for filling factors $\nu\agt 1/6$, only one
peak remains.
The radiation damping contribution
to the linewidth was calculated and found comparable to
experimental linewidths for $\nu\agt 1/4$.
Finally, a study of the effects of
magnetophonon scattering showed that it
has small effects on the spectrum,
except under special circumstances.
All of the results are in good agreement with experiments.

I have benefited from interesting discussions with
Bo Persson on several occasions.
I thank Alan Luther for useful comments on the manuscript.

\begin{figure}
\caption{Calculated absorption spectra. (a) The electron density
is varied as indicated next to the curves, while the magnetic field
and temperature are kept constant. The minority spin
concentration is 10 \%.
At the lowest density, the two spin species give one
resonance peak each, but anti-screening effects transfer
spectral strength from the high-frequency, to the
low-frequency peak.
Eventually, with increasing density, {\em one} motionally
narrowed peak emerges. It starts to broaden at the highest
densities because of radiation damping.
(b) Spectra for two different electron densities
$n_{e}$ (3.8$\times10^{10}$ cm$^{-2}$
and 2.7$\times10^{10}$ cm$^{-2}$),
with varying
temperature (and minority spin concentration) as indicated
next to the curves. The spectra split in qualitatively different
ways depending on $n_{e}$.}
\label{Fig1}
\end{figure}

\begin{figure}
\caption{Absorption spectra calculated
with and without exciton-magnetophonon (MP) scattering.
The MP's are not effective in
scattering long-wavelength excitons.
Thus, for the higher electron density, only
the high-frequency peak is considerably affected by
the MP's.
At the lower density, both peaks are slightly broadened.}
\label{Fig2}
\end{figure}
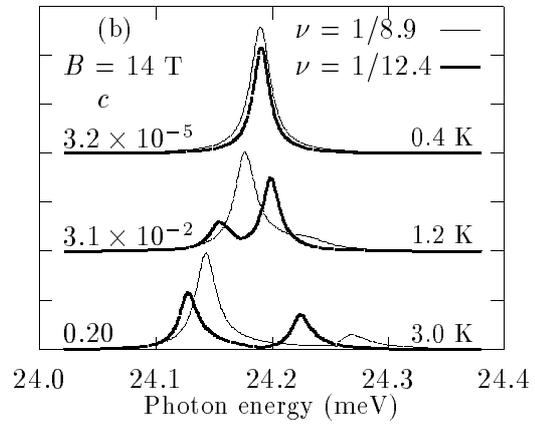
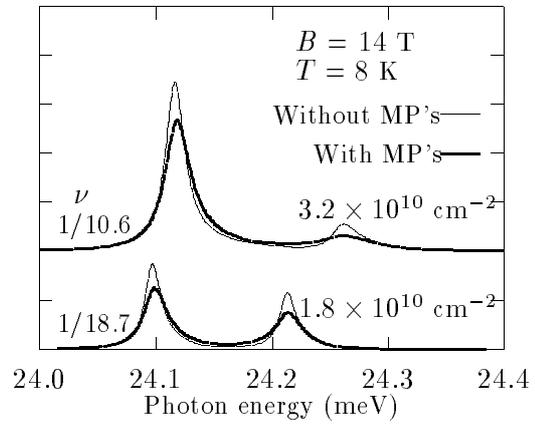
\end{document}